\documentclass[final,3p,times,onecolumn]{elsarticle}
\usepackage{lineno,hyperref}

\usepackage{circledsteps}

\usepackage{amsmath}
\usepackage{graphicx,epsfig,wrapfig}
\usepackage[caption=false]{subfig}

\usepackage{amsthm,mathrsfs}
\usepackage{makeidx}
\usepackage{amsfonts}
\usepackage{amssymb}
\usepackage{amsmath}
\usepackage{mathtools}
\usepackage{fixmath}
\usepackage{amsbsy}
\usepackage[capitalize]{cleveref}
\usepackage{mathrsfs}
\usepackage{slashed}
\usepackage{bm}
\usepackage{multirow}
\usepackage{booktabs}
\usepackage{adjustbox}
\usepackage{comment}

\usepackage{color}
\usepackage[normalem]{ulem}

\usepackage[absolute]{textpos}
\textblockorigin{5mm}{5mm}

\definecolor{myGreen}{rgb}{0.2,0.72,0.2}
\definecolor{myMauve}{rgb}{0.58,0,0.82}
\renewcommand\sout{\bgroup \color[rgb]{0.55,0.00,0.99} \ULdepth=-.5ex \ULset}

\usepackage{simplewick}

\usepackage{multirow}
\usepackage{braket}

\newcommand{\polini}{\epsilon}
\newcommand{\polfin}{\epsilon'^*}

\renewcommand{\[}{\begin{equation}}
\renewcommand{\]}{\end{equation}}

\definecolor{darkgreen}{RGB}{0,120,0}

\definecolor{orange}{RGB}{255,165,0}

\allowdisplaybreaks

\modulolinenumbers[5]
\journal{Physics Letters B}
\biboptions{sort&compress}
\begin{document}


\begin{frontmatter}

\title{Perturbative results for the matrix elements of the vector current \\
and the role of different infrared regulators}


\author[ACAaddr,BPaddr]{Alessio Carmelo Alvaro}
\ead{alessiocarmelo.alvaro01@universitadipavia.it}
\author[JCaddr]{Ignacio Castelli}
\ead{jorge.castelli@temple.edu}
\author[CLaddr]{C\'edric Lorc\'e}
\ead{cedric.lorce@polytechnique.edu}
\author[JCaddr]{Andreas Metz}
\ead{metza@temple.edu}
\author[ACAaddr,BPaddr]{Barbara Pasquini}
\ead{barbara.pasquini@unipv.it}
\author[ACAaddr]{Simone Rodini}
\ead{simone.rodini@unipv.it}

\address[ACAaddr]{Dipartimento di Fisica ``A. Volta", Universit\`a degli Studi di Pavia, 27100 Pavia, Italy}
\address[BPaddr]{Istituto Nazionale di Fisica Nucleare, Sezione di Pavia, 27100 Pavia, Italy}
\address[JCaddr]{Department of Physics, SERC, Temple University, Philadelphia, PA 19122, USA}
\address[CLaddr]{CPHT, CNRS, \'Ecole polytechnique, Institut Polytechnique de Paris, 91120 Palaiseau, France}

\begin{abstract} 
We investigate the twist-2 unpolarized generalized parton distributions (GPDs) of quarks for an on-shell gluon target in quantum chromodynamics. 
These GPDs parametrize the leading-twist matrix elements of the nonlocal light-like flavor-singlet vector current.  
We compute them at one-loop accuracy in perturbation theory using a quark mass and dimensional regularization as infrared regulators.
In particular, we discuss the limit of vanishing momentum transfer.
The present work extends our previous related study on the axial current.
\end{abstract}

\date{\today}


\end{frontmatter}

\section{Introduction}
\label{s:introduction}
Factorization theorems in quantum chromodynamics (QCD)~\cite{Collins:1989gx} are the backbone of the description of hard scattering processes such as inclusive deep-inelastic lepton-nucleon scattering (DIS) and deeply virtual Compton scattering (DVCS)~\cite{Ji:1996ek, Radyushkin:1996nd, Ji:1998xh, Collins:1998be}. 
They allow the cross section for these reactions to be expressed as a convolution of perturbatively calculable short-distance partonic processes and nonperturbative matrix elements that encode information about the partonic structure of the nucleon. 
For DIS and DVCS, the latter are parametrized in terms of parton distribution functions (PDFs) and generalized parton distribution functions (GPDs), respectively~\cite{Muller:1994ses, Ji:1996ek, Ji_1997, Radyushkin:1996nd, Radyushkin:1996ru, Lorce:2025aqp}.

The short-distance processes are typically calculated for vanishing transverse momentum transfer (forward scattering) and using dimensional regularization (DR) to handle the infrared (IR) singularities that can arise from collinear and soft parton kinematics. 
More recently, it has been suggested that a nonvanishing momentum transfer could also be used as an IR regulator for both DIS and DVCS~\cite{Tarasov:2020cwl, Tarasov:2021yll, Bhattacharya:2022xxw, Bhattacharya:2023wvy}.
A special focus was on the photon-gluon fusion process which was computed for off-forward kinematics at one loop, that is, the lowest nontrivial order in the strong coupling $\alpha_{\rm s}$ for this reaction. 
(In DIS the imaginary part of the amplitude for $\gamma^\ast g \! \to \! \gamma^\ast g$ contributes, with $\gamma^\ast$ representing a virtual photon, while in DVCS both the real and imaginary parts of the amplitude for $\gamma^\ast g \! \to \! \gamma g$ contribute.)
It was found that going to off-forward kinematics gives rise to new contributions, extending earlier work related to the proton spin crisis~\cite{Altarelli:1988nr, Carlitz:1988ab, Jaffe:1989jz, Bodwin:1989nz, Anselmino:1994gn, Lampe:1998eu}. 
In particular, it was argued that these additional terms diverge in the forward limit, and potential implications for factorization theorems in DIS and DVCS were discussed~\cite{Tarasov:2020cwl, Tarasov:2021yll, Bhattacharya:2022xxw, Bhattacharya:2023wvy}. 

Another way to explore these new contributions, and especially their behavior in the forward limit, is to compute the matrix elements of the twist-2 light-like flavor-singlet axial and vector currents for a gluon target. 
These matrix elements are parametrized in terms of quark GPDs for a gluon.
In a previous work~\cite{Castelli:2024eza}, we performed such a study for the axial current using the quark mass as an IR regulator. 
For on-shell gluons, we found an additional term (GPD) for off-forward kinematics, in agreement with Refs.~\cite{Tarasov:2020cwl, Tarasov:2021yll, Bhattacharya:2022xxw, Bhattacharya:2023wvy}. 
We also confirmed the important result that this extra term is directly related to the (local) axial anomaly~\cite{Jaffe:1989jz, Tarasov:2020cwl, Tarasov:2021yll, Bhattacharya:2022xxw, Bhattacharya:2023wvy}.
(The nonlocal axial anomaly was studied in Refs.~\cite{Mueller:1997zu, Agaev:2014wna, Bhattacharya:2024geo}.)
However, when a quark mass is used as the IR regulator, together with physical polarization vectors for the external gluons, this additional contribution is not singular in the forward limit and, in fact, vanishes~\cite{Castelli:2024eza}. 
This outcome, which was confirmed in Ref.~\cite{Tarasov:2025mvn} and which generalizes the corresponding result for the local axial current (see~\cite{Anselmino:1994gn, Adler:2004qt, Coriano:2014gja, Castelli:2024eza} and references therein), is required by the conservation of angular momentum~~\cite{Castelli:2024eza}.

The present work extends our previous analysis~\cite{Castelli:2024eza} to the matrix elements of the nonlocal vector current for a gluon target. 
In Sec.~\ref{s:definitions}, we present the decomposition of these matrix elements in terms of (two) unpolarized quark GPDs. 
We also discuss the conditions that these GPDs must satisfy in the forward limit.
Section~\ref{s:nonzero_mass} contains our GPD results for a nonzero quark mass. 
As expected from angular momentum conservation, we find that the contribution of one of the GPDs vanishes in the forward limit. Section~\ref{s:dim_reg} addresses the corresponding calculation using DR, yielding results that are in general agreement with those obtained using the quark-mass regulator.
We conclude in Sec.~\ref{s:conclusions}. 
In~\ref{a:general-results}, we provide general expressions for the two GPDs, from which the results presented in Secs.~\ref{s:nonzero_mass} and~\ref{s:dim_reg} can be derived. 
Whenever applicable, we compare our results with those in the literature.

\section{Unpolarized parton distributions}
\label{s:definitions}
The forward matrix elements of the leading-twist component of the light-like vector current between on-shell gluons  are\footnote{To keep the notation simple, we do not explicitly sum over quark flavors, although this is, strictly speaking, required when considering the flavor-singlet current. None of our main conclusions depends on this point. We also suppress the renormalization scale in the arguments of the parton correlators, PDFs, and GPDs.} 
\begin{equation}
    \Phi_{\lambda \lambda'}^{[\gamma^+]}(x) = \int \frac{dz^-}{4\pi} \, e^{i \, k \, \cdot \, z } \, \langle g (p, \lambda') \, | \, \bar{q}(- \tfrac{z}{2}) \, \gamma^+  \, {\cal W}(-  \tfrac{z}{2}, \tfrac{z}{2}) \, q(\tfrac{z}{2}) \, | \, g (p, \lambda) \rangle \big|_{z^+ = \, 0, \, \vec{z}_\perp = \, \vec{0}_\perp} = - \, ( \polini \cdot \polfin ) \, f_1^{\,q}(x) \,,
    \label{e:forward-matrix-element}
\end{equation}
where $p$ is the momentum of the external gluons and  ${\cal W} (z_1, z_2)$ is the straight Wilson line connecting the points $z_1$ and $z_2$.   The polarization vectors of the initial and final  gluons are denoted by $\polini$ and $\polfin$, respectively, with the dependence on $\lambda$ and $\lambda'$ understood.
The light-front components of a four-vector $a$ are defined as $a^{\,\mu}\!=\!(a^+,a^-,\vec a_\perp)\,$, with $a^{\, \pm}\!=\!(a^0\pm a^3)/\!\sqrt{2}\,$.
Generally, the twist-2 vector current describes unpolarized quarks.
The (dimensionless) function $f_1^{\,q}(x)$ is the unpolarized PDF, which is given by the following linear combination of matrix elements,
\begin{equation}
    f_1^{\,q}(x) = \frac{1}{2} \left( \Phi_{++}^{[\gamma^+]}(x) + \Phi_{--}^{[\gamma^+]}(x) \right) \,.
    \label{e:f1_def}
\end{equation}
This function may show a discontinuity at $x=0$ due to its odd C-parity, $f_1^{\,q}(-x)=-\,f_1^{\,q}(x)\,$.

The matrix elements with different momenta $p$ and $p'$ are parametrized according to
\begin{align}
\label{e:decomposition-GPD}
    F_{\lambda \lambda'}^{[\gamma^+]}(x,\Delta) &= \int \frac{dz^-}{4\pi} \, e^{i \,k \, \cdot \, z } \, \langle g (p', \lambda') \, | \, \bar{q}(- \tfrac{z}{2}) \, \gamma^+  \, {\cal W}(- \tfrac{z}{2}, \tfrac{z}{2}) \, q(\tfrac{z}{2}) \, | \, g (p, \lambda) \rangle \big|_{z^+ =\, 0,\, \vec{z}_\perp = \,\vec{0}_\perp} \nonumber\\
    &= S\!_1 \, H_1^q(x,\xi, \Delta^2) + S\!_2 \, H_2^q(x,\xi,\Delta^2) \,,
\end{align}
where 
\begin{align}
S\!_1 = \, & -  (\polini \cdot \polfin) + \frac{\Delta^2}{2\, (P^+)^2} \, \frac{\polini^+ \, (\polfin )^+}{1-\xi^2} - \frac{1}{1+\xi} \, \frac{\polini^+ \, (\polfin \cdot \Delta)}{P^+} + \frac{1}{1-\xi} \, \frac{(\polini \cdot \Delta) \, (\polfin)^+}{P^+} \,, 
    \label{e:S1} \\
    S\!_2 = \, &  2 \, \frac{(\polini \cdot \Delta) \, (\polfin \cdot \Delta)}{\Delta^2} - (\polini \cdot \polfin)\,, 
    \label{e:S2}
\end{align}
with $P\!=\!(p'+p)/2$ the average momentum, $\,\Delta\!=\!p'-p$ the momentum transfer, and $\xi\! =\! - \Delta^+ / (2 P^+)$ the skewness variable.
Here $H_1^q$ and $H_2^q$ are twist-2 unpolarized GPDs. 
The GPD decomposition in Eq.~\eqref{e:decomposition-GPD} was obtained by imposing the Ward identity for both the initial and final gluon, as well as $\epsilon \cdot p\! =\! \epsilon' \cdot p' \!=\! 0\,$.
It was already found in Ref.~\cite{Bhattacharya:2023wvy} that the matrix elements in Eq.~\eqref{e:decomposition-GPD} are parametrized by two GPDs.
(In that paper, the structure multiplying $H_1^q$ holds in the light-front gauge, while $S\!_1$ in Eq.~\eqref{e:S1} is gauge invariant.) 

For our analysis we choose the symmetric frame, in which we have
\begin{align}
P^{\,\mu} = \Bigg( P^+, \, \frac{\vec{\Delta}_\perp^{\, 2}}{8\, (1 - \xi^2) P^+}, \, \vec{0}_\perp \Bigg) \,, \qquad
\Delta^{\,\mu} = \Bigg( -2\, \xi P^+, \, \frac{\xi \, \vec{\Delta}_\perp^{\, 2}}{4\, (1 - \xi^2) P^+}, \, \vec{\Delta}_\perp \Bigg) \,,
\label{e:sym_frame}
\end{align}
with the frame-dependent relation $\Delta^2\!=\!-\vec{\Delta}_\perp^{\,2}/(1-\xi^2)$ and the frame-independent constraints $P\,  \cdot\, \Delta=0$ and $P^2=-\Delta^2/4\,$. 
In this frame, the general form of the initial-state polarization vectors $\polini_{(1)}\,$, $\polini_{(2)}$ in the light-front gauge is
\begin{equation}
    \epsilon^{\,\mu}_{(i)} = \Bigg( 0\, , \, -\, \frac{\vec{\epsilon}_{\perp(i)} \cdot \vec{\Delta}_\perp }{2\,(1+\xi)P^+} , \, \vec{\epsilon}_{\perp(i)} \Bigg) \, ,
\label{e:general-pol-vectors}
\end{equation}
where $\{\vec{\epsilon}_{\perp(i)}\}$ is a right-handed orthonormal basis in the transverse plane, and the minus component is fixed by employing the condition $\epsilon_{(i)} \cdot p = 0\,$.
The final-state polarization vectors are obtained by flipping the sign of both $\xi$ and $\vec\Delta_\perp$. Polarization vectors with definite light-front helicity correspond to $\epsilon_{\pm} = \mp\, (\epsilon_{(1)} \pm i\,\epsilon_{(2)})/\!\sqrt{2}\,$.

Each one of the GPDs corresponds to a specific helicity combination of the incoming and outgoing gluons, 
\begin{equation}
    H_1^q(x,\xi,\Delta^2) = \frac{1}{2} \left( F_{++}^{[\gamma^+]} (x,\Delta) + F_{--}^{[\gamma^+]} (x,\Delta) \right) \,, \quad H_2^q(x,\xi,\Delta^2) = \frac{1}{2} \left(e^{-2\,i \, \varphi}\, F_{+-}^{[\gamma^+]} (x,\Delta) + e^{2\,i \, \varphi}\, F_{-+}^{[\gamma^+]} (x,\Delta) \right) \,.
    \label{e:GODs-amplitudes}
\end{equation}
Here the angle $\varphi$ is defined as $\cos\varphi = \vec{\epsilon}_{\perp(1)} \cdot  \vec{\Delta}_\perp \, / \, | \,\vec{\Delta}_\perp |\,$, and $\varphi\!=\!\pi$ corresponds to our choice in Ref.~\cite{Castelli:2024eza}. 
From the expressions in Eq.~\eqref{e:GODs-amplitudes}, we identify $H_1^q$ as the helicity-conserving GPD and $H_2^q$ as the helicity-flip GPD.
In the forward limit, the amplitude in Eq.~\eqref{e:decomposition-GPD} must reduce to the amplitude in Eq.~\eqref{e:forward-matrix-element}, which implies 
\begin{equation}
    \lim_{\Delta\to \, 0} \big(S\!_1 \, H_1^q \big) = -\,(\polini\cdot\polfin) \, f_1^{\,q} \, , \qquad \lim_{\Delta\to \, 0} \big( S\!_2 \, H_2^q \big) = 0 \,.
    \label{e:GPDtoPDFcondition}
\end{equation}
From the first condition we see that, in the forward limit, the helicity-conserving GPD reduces to the PDF.
The second condition follows from the conservation of angular momentum, which forbids a helicity-flip transition in the limit $\Delta^2\! \to\! 0\,$. 
Since $S\!_2\,|_{\vec{\Delta}_\perp=\,\vec{0}_\perp}\!\ne\!0\,$ for helicity-flip transitions, that condition can be verified only if $H_2^q$ vanishes in that limit.
The GPDs must therefore satisfy
\begin{equation}
    \lim_{\Delta\to\, 0} \,  H_1^q(x,\xi,\Delta^2)= f_1^{\,q}(x) \, , \qquad \lim_{\Delta^2\to \, 0} H_2^q (x,\xi,\Delta^2) = 0\, .
    \label{e:GPD_forward}
\end{equation} 
As we discuss below, our perturbative results for both $H_1^q$ and $H_2^q$ indeed satisfy the constraints in Eq.~\eqref{e:GPD_forward}. 
An important point raised in Refs.~\cite{Bhattacharya:2022xxw, Bhattacharya:2023wvy} concerns the $\Delta^2$ term in the denominator of $S\!_2\,$.
Specifically, it was argued that this factor would give rise to a singularity when taking the limit $\Delta^2\!\to\! 0\,$. 
This would further imply that computing the matrix elements of the vector current for off-forward kinematics and subsequently taking the forward limit would not agree with a calculation performed directly in forward kinematics.
In contrast, we find that $S\!_2$ is not singular in the limit $\Delta^2\!\to\!0$ and the GPD $H_2^q$ must vanish; consequently, the potentially problematic term vanishes in that limit. 
The discussion in this paragraph also applies to the axial current studied previously~\cite{Tarasov:2020cwl, Tarasov:2021yll, Bhattacharya:2022xxw, Bhattacharya:2023wvy, Castelli:2024eza}.

\section{One-loop results with a massive quark}
\label{s:nonzero_mass}

\begin{figure}[t]
\begin{center}
\includegraphics[width = 0.25 \textwidth]{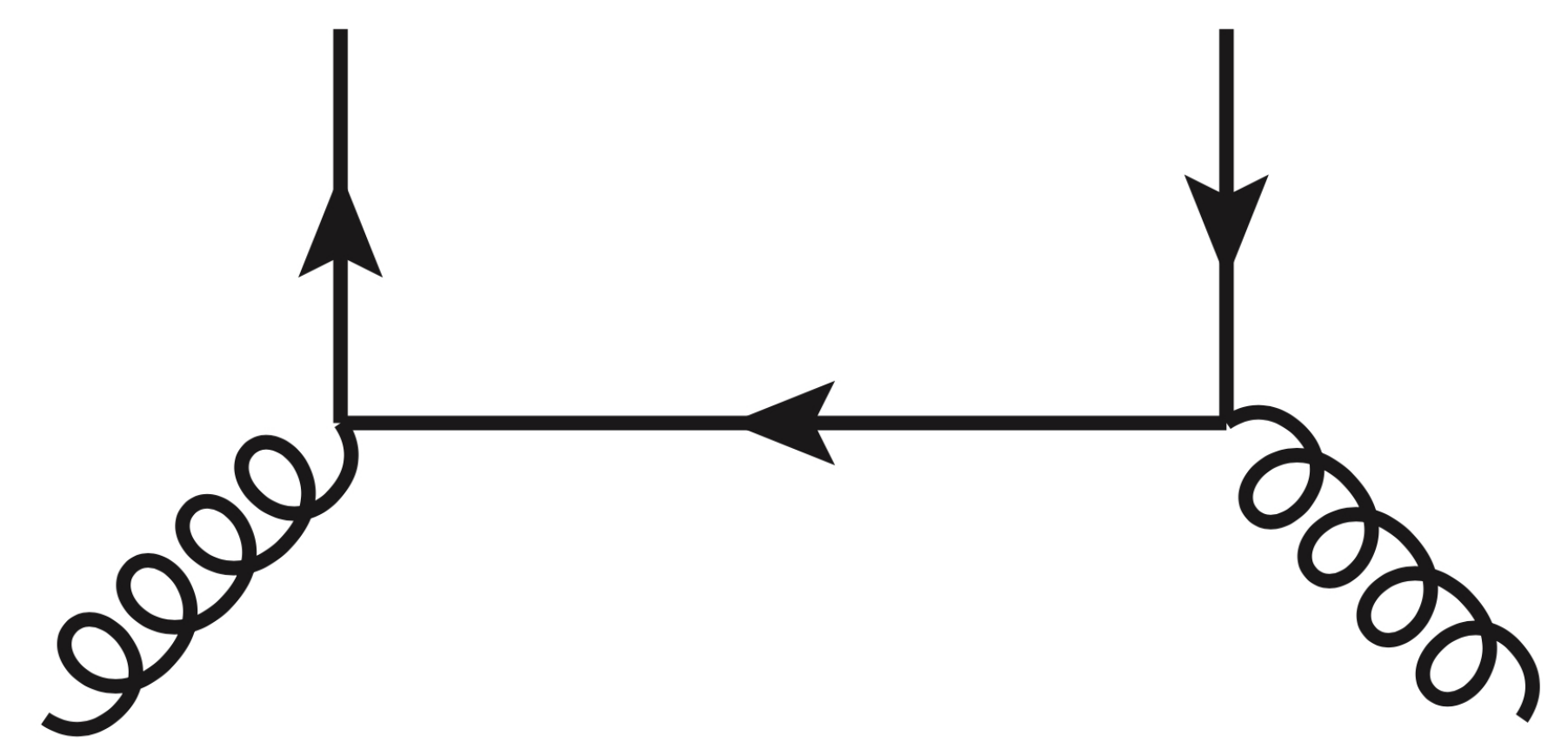} \end{center}
\vspace{-0.4cm}
\caption{Lowest-order diagram contributing to the PDF in the positive-$x$ region and to the GPDs in the positive DGLAP region ($\xi\!\le\! x\!\le\! 1$) and the ERBL region ($-\,\xi \!<\! x \!< \!\xi$). 
A second diagram with reversed arrows is not shown. It contributes to the negative-$x$ region of the PDF and to the ERBL and negative DGLAP regions for the GPDs.
} 
\label{f:GPD_Diagram}
\end{figure}

The one-loop expressions for the PDF and the GPDs are derived by evaluating the diagram in Fig.~\ref{f:GPD_Diagram}.
In this section, a quark mass serves as IR regulator.
To deal with UV divergences, we use DR in $n\! =\! 4 - 2\, \varepsilon$ spacetime dimensions. 
The result for the PDF reads
\begin{equation}
    f_1^{\,q}(x;m) = \frac{\alpha_{\rm s}}{4\pi} \left( \frac{1}{\varepsilon_{\rm UV}} - \log \frac{m^2}{\Bar{\mu}^2} \right) \begin{dcases}
        \,  (1-x)^2 + x^2   \quad & \phantom{-} 0<x\leq 1 \,, \\[0.2cm]
        \, - \,  \left((1+x)^2 + x^2 \right)  & -1\leq x<0 \,,
    \end{dcases}
    \label{e:unpolarized-PDF}
\end{equation}
where $\bar{\mu}^2 = 4\pi e^{-\,\gamma_E}\mu^2$, with $\mu$ being the DR scale and $\gamma_E$ the Euler-Mascheroni constant. The PDF $f_1^{\,q}$ is UV divergent, as reflected by the presence of the $1/\varepsilon_{\rm UV}$ pole, and is discontinuous at $x=0\,$:
\begin{equation}\label{discontinuity}
    \lim_{x\to \,0^+} f_1^{\,q}(x;m) - \lim_{x\to\, 0^-} f_1^{\,q}(x;m) = \frac{\alpha_{\rm s}}{2\pi} \left( \frac{1}{\varepsilon_{\rm UV}} - \log \frac{m^2}{\Bar{\mu}^2} \right) \,.
\end{equation}

The GPDs are obtained by evaluating the linear combinations in Eq.~\eqref{e:GODs-amplitudes}. 
The general results in $n$ dimensions are given in~\ref{a:general-results}. 
After expanding in powers of $\varepsilon$ we find
\begin{align}
H_1^q(x, \xi, \Delta^2; m)  &= \frac{\alpha_{\rm s}}{4 \pi}
\begin{dcases}
\frac{(1-x)^2 + x^2 -\xi^2}{1-\xi^2}\left( \frac{1}{\varepsilon_{\rm UV}} - \log\frac{m^2}{\Bar{\mu}^2} \right)  -\frac{2x\,(1-x)}{1-\xi^2}  \\
- \,\frac{((1-x)^2 + x^2 -\xi^2)\, \kappa-4x\,(1-x)}{(1-\xi^2)\sqrt{\kappa\,(4+\kappa)}} \log\frac{\sqrt{4+\kappa}+\!\sqrt{\kappa}}{\sqrt{4+\kappa}-\!\sqrt{\kappa}} \quad & \phantom{-} \xi\le x\le 1 \, , \\[0.3cm]
\frac{(x+\xi)\,(\xi-2x+1)}{2\,\xi \,(1+\xi)} \left( \frac{1}{\varepsilon_{\rm UV}} - \log\frac{m^2}{\Bar{\mu}^2} \right) - \, \frac{x}{\xi}\,\frac{x+\xi}{1+\xi} \\
- \, \frac{x}{\xi}\,\frac{1+\xi^2-2x}{2\,(1-\xi^2)} \log\left( 1 + \frac{(\xi^2-x^2)\,(1-\xi^2)}{4\,\xi^2(1-x)^2}\,\kappa \right) \\
-\, \frac{((1-x)^2 + x^2 -\xi^2)\, \kappa-4x\,(1-x)}{2(1-\xi^2)\sqrt{\kappa\,(4+\kappa)}} \log\frac{h_+}{h_-} - (x\to -x)  &-\,\xi< x < \xi \, ,
\end{dcases}
\label{e:H1_unpol_general}
\\[0.2cm]
H^q_2(x, \xi, \Delta^2; m)  &= -\, \frac{\alpha_{\rm s}}{4 \pi}
\begin{dcases}
\frac{2x\,(1-x)}{1-\xi^2} \Bigg( 1 - \frac{2}{\sqrt{\kappa\,(4+\kappa)}}\log\frac{\sqrt{4+\kappa}+\!\sqrt{\kappa}}{\sqrt{4+\kappa}-\!\sqrt{\kappa}}\, \Bigg)  \quad &\phantom{-} \xi\le x\le 1 \, , \\[0.3cm]
\frac{x}{\xi}\,\frac{x+\xi}{1+\xi} - \frac{2x\,(1-x)}{1-\xi^2} \frac{1}{\sqrt{\kappa\,(4+\kappa)}} \log\frac{h_+}{h_-} - (x\to-x) &-\,\xi< x < \xi \, , 
\end{dcases}
\label{e:H2_unpol_general}
\end{align}
where $\kappa=\tau\,(1-x)^2/(1-\xi^2)\,$, $\tau=-\Delta^2/m^2$, and 
\begin{equation}
     h_{\pm} = 4\,\xi \,(1-x) \pm (1-\xi)\,(x+\xi)\sqrt{\kappa} \,\left( \sqrt{4+\kappa} \pm \! \sqrt{\kappa} \, \right)\, .
\end{equation}
The results for the negative DGLAP region ($-1\!\le \! x \! \le\! - \,\xi$) are obtained by substituting $x\!\to\! -x$ in the expressions for the positive DGLAP region and then changing the overall sign. 
We have verified that the GPDs are continuous at the crossover points $x\! =\! \pm \, \xi\,$.
Note that only $H_1^q$ is divergent, both in the UV region and the IR region. 
We also point out that the UV divergent part of the GPD $H_1^q$ agrees with the known quark-gluon mixing evolution kernel~ \cite{Muller:1994ses,Ji:1996nm,Radyushkin:1997ki,Balitsky:1997mj,Radyushkin:1998es,Blumlein:1997pi,Blumlein:1999sc,Bertone:2022frx,Bertone:2023jeh}.

As a next step,  we expand the results in Eqs.~\eqref{e:H1_unpol_general} and \eqref{e:H2_unpol_general} around $\tau\!=\! 0$ (corresponding to finite mass $m$ and $\Delta^2 = 0\,$) and for $\tau\!\to\!\infty$ (corresponding to finite $\Delta^2$ and vanishing quark mass). 
In the limit $\tau\!\to\!0\,$, we obtain 
\begin{align}
H^q_1 (x, \xi, \Delta^2; m) &= \frac{\alpha_{\rm s}}{4 \pi} 
\begin{dcases}
\frac{(1-x)^2 + x^2 -\xi^2}{1-\xi^2} \left( \frac{1}{\varepsilon_{\rm UV}} - \log\frac{m^2}{\Bar{\mu}^2} \right) + \mathcal O(\tau) \quad &\phantom{-}\xi \le x \le 1 \, , \\[0.3cm]
\frac{x}{\xi}\,
\frac{1-\xi}{1+\xi} \left( \frac{1}{\varepsilon_{\rm UV}} - \log\frac{m^2}{\Bar{\mu}^2} \right)+ \mathcal O(\tau) \, &-\,\xi < x < \xi \, ,
\end{dcases} 
\label{e:H1_tau_0}
\\[0.2cm]
H^q_2\,(x, \xi, \Delta^2; m) &= -\, \frac{\alpha_{\rm s}}{4 \pi} 
\begin{dcases}
-\, \frac{x\,(1-x)^3}{3\,(1-\xi^2)^2} \,\tau + \mathcal O(\tau^2) \overset{\tau\to \,0}{\to} 0 \quad &\phantom{-}\xi \le x \le 1 \, , \\[0.3cm]\frac{x}{\xi}\,
\frac{3x^2-\xi \,(2+\xi)}{6 \,(1+\xi)^2}\,\tau + \mathcal O(\tau^2) \overset{\tau\to \, 0}{\to} 0 \, &-\,\xi < x < \xi \, .
\end{dcases}
\label{e:H2_tau_infty}
\end{align}
For $\xi\!=\! 0\,$, the result for the helicity-conserving GPD $H_1^q$ in Eq.~\eqref{e:H1_tau_0} reduces to that of the PDF in Eq.~\eqref{e:unpolarized-PDF}.
Furthermore, the helicity-flip GPD $H^q_2$  in Eq.~\eqref{e:H2_tau_infty}  vanishes in the limit $\Delta^2\!\to\! 0\,$.
Our results for the two GPDs therefore satisfy the constraints in Eq.~\eqref{e:GPD_forward}\,.

In the limit $\tau\!\to\! \infty\,$, we find instead
\begin{align}
\label{e:H1-BHV}
H^q_1 (x, \xi, \Delta^2; m) &= \frac{\alpha_{\rm s}}{4 \pi} 
\begin{dcases}
\frac{(1-x)^2 + x^2 -\xi^2}{1-\xi^2} \left( \frac{1}{\varepsilon_{\rm UV}} - \log\frac{-\Delta^2}{\Bar{\mu}^2} \right) \\
-\, \frac{2x\,(1-x)}{1-\xi^2} - \frac{ (1-x)^2 + x^2 -\xi^2}{1-\xi^2}\log\frac{(1-x)^2}{1-\xi^2} +\mathcal O (\tau^{-1} \log\tau) \quad & \phantom{-}\xi \le x \le 1 \, , \\[0.3cm] 
\
\frac{x}{\xi} \,\frac{1-\xi}{1+\xi} \left( \frac{1}{\varepsilon_{\rm UV}} - \log\frac{-\Delta^2}{\Bar{\mu}^2} \right)  \\
-\,\frac{2x}{1+\xi} + \frac{2x}{1-\xi^2}\log\frac{1-x^2}{(1+\xi)^2} -\, \frac{x}{\xi}\,\frac{1+\xi^2}{1-\xi^2} \log\frac{\xi^2-x^2}{4\,\xi^2} \\
- \,\frac{2x^2+1-\xi^2}{1-\xi^2} \log\frac{(1-x)\,(\xi+x)}{(1+x)\,(\xi-x)} + \mathcal O ( \tau^{-1} \log\tau )  &-\,\xi < x < \xi \, ,
\end{dcases}
\\[0.2cm]
\label{e:H2-BHV}
H^q_2\,(x, \xi, \Delta^2; m) &= -\, \frac{\alpha_{\rm s}}{4 \pi} 
\begin{dcases}
\frac{2x\,(1-x)}{1-\xi^2} +\mathcal O ( \tau^{-1} ) \quad & \phantom{-}\xi \le x \le 1 \, , \\[0.3cm]
\frac{2x}{1+\xi} +\mathcal O ( \tau^{-1} )  &-\,\xi < x < \xi \, .
\end{dcases}
\end{align}
These expressions match with the $\varepsilon$-expansion of the results in Ref.~\cite{Bhattacharya:2023wvy}, where the calculations were performed with the quark mass set to zero from the beginning.
We also note that the result in Eq.~\eqref{e:H2-BHV} agrees with the matching coefficient for generalized transverse-momentum-dependent parton distributions presented in Ref.~\cite{Bertone:2025vgy} for what is referred to in that paper as the transition from a linearly polarized gluon to an unpolarized quark.\footnote{
To convert to the notation of Ref.~\cite{Bertone:2025vgy}, one should divide Eq.~\eqref{e:H2-BHV} by the factor $(1-\xi^2)$ and perform the substitutions $x\!\to\! y\,$, $\xi\! \to\! \kappa\, y \,$. Then, the result for the DGLAP region is the top right entry of Table 1 of Ref.~\cite{Bertone:2025vgy}, while the result for the ERBL region should be compared with the ERBL combination of Eq. (23) in Ref.~\cite{Bertone:2025vgy}, whose ingredients can be read off from Tables 4 and 5 of that reference.} 

The terms proportional to $\log(-\Delta^2 / \bar{\mu}^2)$ in Eq.~\eqref{e:H1-BHV} reflect that a nonzero momentum transfer acts as the IR regulator after the $m\!\to\! 0$ limit was taken, and as a consequence, the limit $\Delta^2\!\to\! 0$ is not justified.
However, irrespective of the divergence that arises in the limit $\Delta^2\!\to\! 0\,$, it must be kept in mind that the result in Eq.~\eqref{e:H1-BHV} has been obtained for finite $\Delta^2$. 
Therefore, with the exception of the UV-divergent part, even the terms in Eq.~\eqref{e:H1-BHV} that do not depend on $\Delta^2$ cannot be considered for $\Delta^2 \!=\! 0\,$.
The same argument applies to the result for the GPD $H_2^q$ in Eq.~\eqref{e:H2-BHV}: although the expression is independent of $\Delta^2$, the result must not be considered for $\Delta^2 \! =\! 0\,$.
In fact, neglecting the assumption underlying Eq.~\eqref{e:H2-BHV}, namely the limit $\tau\! \to\! \infty\,$, leads to a contradiction with the angular momentum constraint in Eq.~\eqref{e:GPD_forward}. 
Overall, we find that the limit $\Delta^2\! \to\! 0$ for both GPDs can be taken only when the quark mass is retained as an (additional) IR regulator.
This general result agrees with the corresponding outcome for the GPDs that parametrize the matrix elements of the nonlocal axial current~\cite{Castelli:2024eza}.  
In the next section we arrive at the same conclusion when using DR as the IR regulator.

\section{One-loop results in dimensional regularization}
\label{s:dim_reg}
To present the results for the PDF and GPDs using DR for both the UV and IR divergences, one needs to isolate the two regions of the scaleless integral (see, for instance, Ref.~\cite{Collins:2011zzd})\footnote{One can write in general $[f(\varepsilon_{\rm UV})-f(\varepsilon_{\rm IR})] / {4\pi}\,$ with $f(\varepsilon)=\sum_{n=-1}^\infty c_n \,\varepsilon^n$, where $c_{-1}\!=\!1$ is uniquely fixed.},
\begin{equation}
\mu^{2\varepsilon}\int 
\frac{d^{\,2-2\, \varepsilon}\vec{k}_\perp }{(2\pi)^{2-2\, \varepsilon}}\,
\frac{1}{ \vec{k}^{\,2}_{\perp}}
=
\frac{1}{4\pi} \left(\frac{1}{\varepsilon_{\rm UV}}-\frac{1}{\varepsilon_{\rm IR}}\right) \,.
\label{e:scaleless_int}
\end{equation}

Calculating the Feynman diagrams contributing to the forward matrix elements in Eq.~\eqref{e:forward-matrix-element} and using Eq.~\eqref{e:scaleless_int}, we obtain the following result for the PDF:
\begin{equation}
     f_1^{\,q}(x;\varepsilon_{\rm IR}) = \frac{\alpha_{\rm s}}{4\pi} \, \frac{1}{1-2\, \varepsilon} \left( \frac{1}{\varepsilon_{\rm UV}} - \frac{1}{\varepsilon_{\rm IR}} \right)\begin{dcases}
        \, (1-x)^2 + x^2  \quad & \phantom{-} 0<x\leq 1 \,, \\[0.2cm]
        \, -\left((1+x)^2 + x^2\right)  & -1\leq x<0 \,.
    \end{dcases}\label{e:PDFdimreg}
\end{equation}
The coefficient of the UV pole governs the renormalization group evolution of the PDF; see~\cite{Collins:2011zzd} and references therein. 
The compact notation $f(\varepsilon)\,(1/\varepsilon_{\rm UV} - 1/\varepsilon_{\rm IR})$ stands for  $(f(\varepsilon_{\rm UV})/\varepsilon_{\rm UV} - f(\varepsilon_{\rm IR})/\varepsilon_{\rm IR})\,$.
In Eq.~\eqref{e:PDFdimreg}, we kept the dependence on the regulators $\varepsilon_{\rm UV}$ and $\varepsilon_{\rm IR}$ unexpanded, and likewise for the GPD results that follow in this section. 
 
In computing the GPDs, we encounter an integral of the form
\begin{align}
\mu^{2\, \varepsilon}\int \frac{d^{\,2-2\, \varepsilon}\vec{k}_\perp }{(2\pi)^{2-2\, \varepsilon}}\,
\frac{1}{ \vec{k}^{\,2}_{\perp}+M^2 } 
&
= \mu^{2\, \varepsilon}\int \frac{d^{\,2-2\, \varepsilon}\vec{k}_{\perp} }{(2\pi)^{2-2\, \varepsilon}}\, \Bigg( \frac{1}{\vec{k}_{\perp}^{\,2}+M^2} - \frac{1}{\vec{k}^{\,2}_{\perp}} \Bigg) + \mu^{2\, \varepsilon}\int \frac{d^{\,2-2\, \varepsilon}\vec{k}_{\perp}}{(2\pi)^{2-2\, \varepsilon}}\, \frac{1}{\vec{k}^{\,2}_{\perp}} 
\notag \\
&= \frac{1}{4\pi} \left[ 
\left( \frac{M^2}{4\pi\mu^2} \right)^{-\varepsilon_{\rm IR}} \Gamma(\varepsilon_{\rm IR}) +
\left( \frac{1}{\varepsilon_{\rm UV}} - \frac{1}{\varepsilon_{\rm IR}} \right) 
\right]
    \, ,
\label{e:scale_int}
\end{align}
where $M$ is a physical mass scale.
The result of the integral in Eq.~\eqref{e:scale_int} can be used both for finite $M$ and in the limit $M\! \to\! 0\,$.
In the former case, we can expand the result in powers of $\varepsilon_{\rm IR}$ and readily recover the well-known result that this integral has only a UV divergence.
In the latter case, since $\varepsilon_{\rm IR}\! <\! 0\,$, the term $\big( M^2/(4 \pi \mu^2) \big)^{- \varepsilon_{\rm IR}}$ vanishes for $M \!\to\! 0\,$, leading to agreement with the result in Eq.~\eqref{e:scaleless_int}. 
We emphasize that, after expanding the result in powers of $\varepsilon_{\rm IR}$ through the constant term, the limit $M \! \to \! 0\,$ can no longer be taken due to the presence of $\log M^2/\bar{\mu}^2$.
This outcome is analogous to the discussion in the last paragraph of Sec.~\ref{s:nonzero_mass}, where the roles of $\Delta^2$ and $m$ are here played by $M^2$ and $\varepsilon_{\rm IR}\,$, respectively.
 
The results for the GPDs at finite $\Delta^2$ and unexpanded in $\varepsilon$ are
\begin{align}
H_1^q(x,\xi,\Delta^2;\varepsilon_{\rm IR}) =&\, \frac{\alpha_{\rm s}}{4\pi} 
\begin{dcases}
\frac{ (1-x)^2 + x^2 -\xi^2}{1-\xi^2}\, \frac{1}{1-2\, \varepsilon}  \left( \frac{1}{\varepsilon_{\rm UV}} - \frac{1}{\varepsilon_{\rm IR}} \right)  \\
- \left( -\, \frac{\Delta^2}{4\pi\mu^2} \frac{(1-x)^2}{1-\xi^2} \right)^{-\varepsilon_{\rm IR}} \frac{\Gamma(1+\varepsilon_{\rm IR})}{1-2\, \varepsilon_{\rm IR}} \,\Tilde{B}(\varepsilon_{\rm IR} , 1; \infty)\,  \left( \frac{x^2-x+1-\xi^2}{1-\xi^2}\left(1-\varepsilon_{\rm IR}\right) -\, \frac{1}{2}  \right)  & \phantom{-} \xi \leq x \leq 1 \, , 
\\[0.3cm]
\frac{(1+\xi-2x)\,(x+\xi)}{2\,\xi \,(1+\xi)}\, \frac{1}{1-2\, \varepsilon} \left( \frac{1}{\varepsilon_{\rm UV}} - \frac{1}{\varepsilon_{\rm IR}} \right) \\
- \left( -\, \frac{\Delta^2}{4\pi\mu^2} \frac{(1-x)^2}{1-\xi^2} \right)^{-\varepsilon_{\rm IR}} \frac{\Gamma(1+\varepsilon_{\rm IR})}{1-2\, \varepsilon_{\rm IR}}\, \Tilde{B}(\varepsilon_{\rm IR} , \zeta; \infty)\,  \left( \frac{x^2-x+1-\xi^2}{1-\xi^2}\left(1-\varepsilon_{\rm IR}\right)-\, \frac{1}{2} \right) \\
+ \left( -\, \frac{\Delta^2}{4\pi\mu^2} \frac{\xi^2-x^2}{4\,\xi^2} \right)^{-\varepsilon_{\rm IR}} \frac{\Gamma(\varepsilon_{\rm IR})}{1-2\, \varepsilon_{\rm IR}} \, \frac{x}{\xi} \left( \frac{1-x}{1-\xi^2}\left(1-\varepsilon_{\rm IR}\right)-\frac{1}{2}\right) - (x\to -x) \, & -\,\xi < x < \xi \, ,
\end{dcases}
\label{e:new-H1-BHV}\\
H_2^q(x,\xi,\Delta^2;\varepsilon_{\rm IR}) =&\, \frac{\alpha_{\rm s}}{4\pi} \begin{dcases}
        \frac{x\,(1-x)}{1-\xi^2} \left( -\, \frac{\Delta^2}{4\pi\mu^2} \frac{(1-x)^2}{1-\xi^2} \right)^{-\varepsilon_{\rm IR}} \frac{\Gamma(1+\varepsilon_{\rm IR})}{1-2\, \varepsilon_{\rm IR}} \, \varepsilon_{\rm IR} \, \Tilde{B}(\varepsilon_{\rm IR} , 1; \infty) \,  & \phantom{-} \xi \leq x \leq 1 \, , \\[0.3cm]
        \frac{x\,(1-x)}{1-\xi^2} \left( -\, \frac{\Delta^2}{4\pi\mu^2} \frac{(1-x)^2}{1-\xi^2} \right)^{-\varepsilon_{\rm IR}} \frac{\Gamma(1+\varepsilon_{\rm IR})}{1-2\, \varepsilon_{\rm IR}}  \,\varepsilon_{\rm IR} \, \Tilde{B}(\varepsilon_{\rm IR} , \zeta; \infty) \\
        -\,\frac{x}{\xi}\, \frac{x-\xi^2}{1-\xi^2} \left( -\, \frac{\Delta^2}{4\pi\mu^2} \frac{\xi^2-x^2}{4\,\xi^2} \right)^{-\varepsilon_{\rm IR}} \frac{\Gamma(1+\varepsilon_{\rm IR})}{1-2\, \varepsilon_{\rm IR}} - (x\to -x) \, & -\,\xi < x < \xi \, ,
    \end{dcases}
    \label{e:new-H2-BHV}
\end{align}
with $\zeta\! =\! (x + \xi)\,(1 - \xi) / (2 \xi\, (1 - x))$ and
the function $\Tilde{B}$ defined in Eq.~\eqref{e:Btilde_full}.
Like in the case of the nonzero quark mass, we have checked that both GPDs in Eqs.~\eqref{e:new-H1-BHV} and~\eqref{e:new-H2-BHV} are continuous at $x\! =\! \pm \, \xi\,$.
Note that all but the terms proportional to $(1/\varepsilon_{\rm UV} - 1/\varepsilon_{\rm IR})$
come with a factor $\left( - \Delta^2 / (4\pi\mu^2)\right)^{-\varepsilon_{\rm IR}}$.

Keeping the results unexpanded in $\varepsilon_{\rm IR}$ allows a safe expansion in powers of $\Delta^2$, providing 
\begin{align}
    H_1^q(x,\xi,\Delta^2;\varepsilon_{IR}) &= \frac{\alpha_{\rm s}}{4\pi} \begin{dcases}
        \frac{ (1-x)^2 + x^2 -\xi^2}{1-\xi^2}\, \frac{1}{1-2\, \varepsilon}  \left( \frac{1}{\varepsilon_{\rm UV}} - \frac{1}{\varepsilon_{\rm IR}} \right) + {\cal O}\left((-\Delta^2 / \mu^2)^{- \varepsilon_{\rm IR}}\right) \quad & \phantom{-} \xi \leq x \leq 1 \, , \\[0.3cm]
        \frac{x}{\xi}\, \frac{1-\xi}{1+\xi}\, \frac{1}{1-2\, \varepsilon} \left( \frac{1}{\varepsilon_{\rm UV}} - \frac{1}{\varepsilon_{\rm IR}} \right) + {\cal O}\left((-\Delta^2 / \mu^2)^{- \varepsilon_{\rm IR}}\right)  & -\,\xi < x < \xi \, ,
    \end{dcases} \label{e:H1_DR_forward}
    \\[0.2cm]
    H_2^q(x,\xi,\Delta^2;\varepsilon_{IR})& = {\cal O}\left((-\Delta^2 / \mu^2)^{- \varepsilon_{\rm IR}}\right)\, .
    \label{e:H2_DR_forward}
\end{align}
For $\xi \!= \!0\,$, the expression for $H_1^q$ in Eq.~\eqref{e:H1_DR_forward} coincides with the result for $f^{\,q}_1$ in Eq.~\eqref{e:PDFdimreg}, and $H_2^q$ in Eq.~\eqref{e:H2_DR_forward} vanishes in the limit $\Delta^2\!\to\! 0\,$.
The results in Eqs.~\eqref{e:H1_DR_forward} and ~\eqref{e:H2_DR_forward} are therefore in agreement with the constraints in Eq.~\eqref{e:GPD_forward}.
Instead, if we first expand Eqs.~\eqref{e:new-H1-BHV} and~\eqref{e:new-H2-BHV} around $\varepsilon_{\rm IR}\!=\!0\,$,
we reproduce the results in Eqs.~\eqref{e:H1-BHV} and~\eqref{e:H2-BHV}, for which the limit $\Delta^2 \! \to \! 0$ cannot be taken, as discussed above.
Overall, our findings in this section are consistent with those in Sec.~\ref{s:nonzero_mass}, in that another IR regulator must be kept in the GPD results to obtain results in the limit $\Delta^2 \!\to\! 0\,$.

\section{Conclusions}
\label{s:conclusions}

In a previous work~\cite{Castelli:2024eza}, we studied, to leading order in perturbative QCD, the matrix elements of the twist-2 nonlocal axial current evaluated between external on-shell gluon states, and addressed the question of its regularity in the limit $\Delta^2 \!\to\! 0\,$. 
We found that using a quark mass as an IR regulator allows this limit to be taken and leads to results consistent with angular momentum conservation.
To investigate whether this conclusion is specific to the axial current and the quark-mass regulator, in the present work we considered the corresponding twist-2 nonlocal vector current, using two alternative IR regularization schemes: a finite quark mass and dimensional regularization.
In agreement with Ref.~\cite{Bhattacharya:2023wvy}, we found that the matrix elements of the vector current can be parametrized in terms of two quark GPDs for a gluon. 
We further showed that one of these GPDs is helicity-conserving, whereas the other one describes a gluon helicity flip that must vanish in the limit $\Delta^2\!\to\! 0$ due to angular momentum conservation.
In analogy with our previous analysis~\cite{Castelli:2024eza}, we found that the limit $\Delta^2\! \to\! 0$ can be taken only if $\Delta^2$ does not act as the sole IR regulator, unless one is only interested in the UV-divergent part. 
On the other hand, when either a quark mass or DR is used as an IR regulator, we find that both GPDs have a well-defined forward limit and are in full agreement with calculations performed directly in forward kinematics.

Our one-loop results are in agreement with the general constraints in Eq.~\eqref{e:GPD_forward}, which we expect to hold to all orders in perturbation theory. Furthermore, we expect them to have implications for the box diagram. 
For instance, it is known that in DVCS a gluon helicity flip in the process $\gamma^\ast g \! \to\! \gamma g$ can occur in forward kinematics only if there is a corresponding helicity flip of the photons~\cite{Hoodbhoy:1998vm, Belitsky:2000jk, Diehl:2001pm}. 
Therefore, if this process is computed for helicity-conserving photon transitions at nonzero $\Delta^2$, the contribution associated with a gluon helicity flip must vanish when taking the limit $\Delta^2 \!\to\! 0\,$.
Based on the results of the present work, exactly this behavior should arise when, in addition to $\Delta^2$, a quark mass, DR, or any other IR regulator is used.
Finally, while our work emphasizes the regularity of the matrix elements in the limit $\Delta^2\!\to\! 0\,$, it does not interfere with other aspects developed in Refs.~\cite{Tarasov:2020cwl, Tarasov:2021yll, Bhattacharya:2022xxw, Bhattacharya:2023wvy}.

\section*{Acknowledgement}
We have used the FeynCalc~\cite{Shtabovenko:2020gxv} and JaxoDraw~\cite{Binosi:2008ig} software.
A.C.A.~acknowledges the hospitality of Temple University, where part of this research was conducted.
I.C.~acknowledges helpful discussions with Farid Salazar and the hospitality of \'Ecole polytechnique and Universit\`a degli Studi di Pavia. 
The work of I.C.~was supported by the National Science Foundation under the Grant No.~PHY-2412792 and by the U.S.~Department of Energy, Office of Science, Office of Nuclear Physics under the Quark-Gluon Tomography (QGT)
Topical Collaboration with Award DE-SC0023646.
The work of A.M.~was supported by the National Science Foundation under the Grant No. PHY-2412792. 

\appendix
\section{One-loop results for a massive quark in $n$ dimensions}
\label{a:general-results}
Here we present the most general results for the GPDs, namely for a nonzero quark mass and in $n\!=\!4-2\,\varepsilon$ spacetime dimensions.
In the DGLAP region, we find
\begin{align}
    H_1^q(x,\xi,\Delta^2) &=\frac{\alpha_{\rm s}}{4\pi} \left[ \left(1-\frac{2x\,(1-x)}{1-\xi^2}\frac{1-\varepsilon}{1-2\, \varepsilon}\,\right) D_1 -\, \frac{4}{\tau}\, \frac{x}{1-x}\,\frac{D_2}{1-2\, \varepsilon} \right] , \\
    H_2^q(x,\xi,\Delta^2) &=-\, \frac{\alpha_{\rm s}}{4\pi}\left[
    \frac{ 2x \, (1-x)}{1-\xi^2}\, \frac{\varepsilon \,D_1}{1-2\,\varepsilon}+\frac{4}{\tau}\, \frac{x}{1-x}\,\frac{D_2}{1-2\, \varepsilon}
    \right]
    \, 
    ,
\end{align}
where 
\begin{align}
    D_1 &=
    \left(\frac{m^2}{4\pi \mu^2}\right)^{-\varepsilon} \Gamma(1+\varepsilon)\left(\frac{-\kappa^{-\varepsilon}}{2}\, \Tilde{B} \left(\varepsilon_{\rm IR} , 1 ; \kappa\right)+\frac{1}{\varepsilon_{\rm UV}}\right)\,,
    \\
    D_2 &=-\,\frac{1}{2}\left(\frac{m^2}{4\pi \mu^2} \right)^{-\varepsilon} \Gamma(1+\varepsilon) \, \kappa^{-\varepsilon}\, \Tilde{B} \left(\varepsilon_{\rm IR} , 1 ;\kappa\right) \,,
\end{align}
and $\kappa$ is given after Eq.~\eqref{e:H2_unpol_general}.
The function $\Tilde{B} \left(\varepsilon, \zeta;\kappa\right)$ is defined as
\begin{gather}
    \Tilde{B} \left(\varepsilon, \zeta;\kappa\right)
    =\int_0^{\,\zeta} d\alpha \, \left(\kappa^{-1}+\alpha\,(1-\alpha)\right)^{-1-\varepsilon} \,.
    \label{e:Btilde_full}
\end{gather}
This function is regular for $\kappa^{-1}\! >\! 0\,$, but is divergent for $\kappa^{-1}\!=\!0$ as $\varepsilon\! \to\! 0\,$.
Defining $\alpha_{\pm}\!=\!\frac{1\pm \sqrt{1+4\,\kappa^{-1} }}{2}\,$, we obtain the expansion
\begin{gather}
    \Tilde{B} \left(\varepsilon, \zeta;\kappa\right) = 
    -\,\frac{\left(\zeta-\alpha_-\right)^{-\varepsilon}+\left(\alpha_+\right)^{-\varepsilon}-\left(\alpha_+-\zeta\right)^{-\varepsilon}-\left(-\alpha_-\right)^{-\varepsilon}}{\left(\alpha_+-\alpha_-\right)^{1+\varepsilon}}\, \frac{1}{\varepsilon}
    +{\cal O} \big( \varepsilon \big) \,,
\end{gather}
and, in particular,
\begin{equation}
\varepsilon \, \Tilde{B} \left(\varepsilon, \zeta;\kappa\right) \stackrel{\varepsilon\,\to\,0}{\longrightarrow} -\,2+{\cal O} \big( \varepsilon \big) \,. 
\end{equation}

The results for the GPDs in the ERBL region read
\begin{align}
    \begin{split}
        H^q_1(x,\xi,\Delta^2) =& \frac{\alpha_{\rm s}}{4\pi}\left[ \left(1-
        \frac{2  x\,(1-x) }{1-\xi^2}\, \frac{1-\varepsilon}{1-2\,\varepsilon}\right)E_1-\frac{4}{\tau}\, \frac{x}{1-x}\, \frac{E_2}{1-2\, \varepsilon} - 2\, \frac{\xi^2-x^2}{1-\xi^2}\, \frac{1-\varepsilon}{1-2\,\varepsilon}\, E_3\,  \right]- (x\to -x)\,, 
    \end{split} \\
        H^q_2(x,\xi,\Delta^2) =&\, -\frac{\alpha_{\rm s}}{4\pi}\,
        \left[\frac{2  x \, (1-x)}{1-\xi^2} \, \frac{\varepsilon \, E_1}{1-2\,\varepsilon}+\frac{4}{\tau}\, \frac{x}{1-x}\, \frac{E_2}{1-2\, \varepsilon}+2\,\frac{ \xi^2-x^2}{1-\xi^2} \, \frac{\,\varepsilon \,E_3}{1-2\,\varepsilon}\right] - (x\to -x)
        \, ,
\end{align}
where 
\begin{align}
    E_1&= \, \left(\frac{m^2}{4\pi \mu^2} \right)^{-\varepsilon}\Gamma(1+\varepsilon) \left[\frac{-\kappa^{-\varepsilon}}{2}\, \Tilde{B} \left(\varepsilon_{\rm IR}, \zeta;\kappa\right)+\frac{1}{2}\left(1+\frac{x}{\xi}\,(1+\zeta\,(1-\zeta) \,\kappa)^{-\varepsilon}\right)\frac{1}{\varepsilon_{\rm UV}}\,\right]\, ,\\
    E_2&= -\, \frac{1}{2}\, \left(\frac{m^2}{4\pi \mu^2} \right)^{-\varepsilon}\Gamma(1+\varepsilon)\, \kappa^{-\varepsilon}\, \Tilde{B} \left(\varepsilon_{\rm IR}, \zeta;\kappa\right) \, ,\\
    \quad E_3&= -\,
    \frac{1}{2}\,\frac{x}{\xi}\, \left(\frac{m^2}{4\pi \mu^2}\right)^{-\varepsilon}\Gamma(1+\varepsilon) \,(1+\zeta\,(1-\zeta) \,\kappa)^{-\varepsilon} \, \frac{1}{\varepsilon_{\rm UV}}\,.
\end{align}
For the results in the ERBL region, the variable $\zeta$ as defined after Eq.~\eqref{e:new-H2-BHV} is to be used.

\bibliography{biblio}

\end{document}